\documentclass[%
 preprint, amsmath,amssymb,
 aps, physrev,
]{revtex4-2}

\usepackage{graphicx}% 
\usepackage{dcolumn}% 
\usepackage{bm}% 
\usepackage{siunitx}
\usepackage{ulem}
\usepackage{nccmath}
\usepackage{cancel}

\begin{document}

\title{\textbf{Analytical dispersion relation for forward volume spin waves in ferrimagnets near the angular momentum compensation condition} 
}% 

\author{Luis S\'{a}nchez-Tejerina}
 \affiliation{Department of Electricity and Electronics, University of Valladolid, Valladolid, Spain.}%
 \email{Contact author: luis.sanchez-tejerina@uva.es}
 
 \author{David Osuna Ruiz}
    \affiliation{Departament of Electrical, Electronical and Communications Engineering, Public University of Navarra, Pamplona, Spain}%
 
\author{V\'{i}ctor Raposo}%
    \affiliation{Department of Applied Physics, University of Salamanca, Salamanca, Spain}%
       
\author{Eduardo Mart\'{i}nez}
    \affiliation{Department of Applied Physics, University of Salamanca, Salamanca, Spain}%

\author{Luis L\'{o}pez Díaz}
    \affiliation{Department of Applied Physics, University of Salamanca, Salamanca, Spain}%

\author{\'{O}scar Alejos}
 \affiliation{Department of Electricity and Electronics, University of Valladolid, Valladolid, Spain.}%
\date{\today}%

\begin{abstract}
Antiferromagnetic magnonics has become the focus of intense scientific research because of the advantages of these materials compared to ferromagnets. However, ferrimagnetic materials have received much less attention despite exhibiting similar dynamical features at the angular momentum compensation point. In this paper, we present analytical expressions describing the dispersion relation of forward volume spin waves in ferrimagnetic materials near the angular momentum compensation point. Besides, we benchmark the derived dispersion relations with full micromagnetic simulations showing a complete agreement between both approaches. This work predicts two different branches for forward volume spin waves in ferrimagnetic materials merging into a single branch at the angular momentum compensation point. Our results can assist in the design of magnonic devices built on ferrimagnetic materials.
\end{abstract}

\maketitle

Magnon propagation in antiferromagnetic (AFM) and/or ferrimagnetic (FiM) materials is being the subject of numerous studies in recent times~\cite{Wang:15,Lisenkov:2019,Rezende:19,Cheng:16,Puliafito:19,Xu:19,Yanes:20,Safin:20,Chen:23,Sanchez-Tejerina:24} because of their particular characteristics in terms of higher frequency dynamics, minimal sensitivity to external fields, and the existence of a wide range of available materials, mostly electrical insulators, all compared to ferromagnetic (FM) materials. Another important aspect is that the spin transport in such materials (a key concept in spintronics) is primarily due to magnons, given the absence of conduction electrons, yet another difference with respect to FM materials, where electrons are the main spin carriers. This fact establishes an immediate connection between magnonics and spintronics. Furthermore, the absence of electron transport minimizes the importance of thermal effects from local heating.

Significant effort has been made to analytically describe the magnetization dynamics of spins in magnetic materials. Kalinikos and Slavin's work on FM materials~\cite{Kalinikos:86} established a precedent for analyzing spin wave dispersion across a broad frequency spectrum. In recent years, there has been a notable increase in theoretical studies of spin dynamics in AFM materials. Among several examples, we can mention the characterization of the fundamental resonance modes $k = 0$~\cite{Rezende:19, Cheng:16, Puliafito:19}, magnon propagation in uniform films~\cite{Xu:19,Millo:23}, or through material interfaces~\cite{Gorobets:2023}, or the interaction of AFM domain walls (DW) and magnons~\cite{Yanes:20, Wang:15}, always focused on their fundamental physics, since knowledge of AFM magnonics is still scarce~\cite{Rezende:19, Safin:20, Barman:21}. Another fact is that FiM materials can mimic the dynamic response of AFMs at the point of angular momentum compensation (AMC)~\cite{Caretta:18, Siddiqui:18, Cutugno:21} with the advantage that they are at the same time easier to manipulate and detect by electrical and/or optical means~\cite{Ueda:2016, Fleischer:18, Hortensius:21}. Indeed, localized precession of spins in FiM materials reaching sub-THz regimes, have been analytically predicted~\cite{Lisenkov:2019}, with applications in high-frequency spintronic devices~\cite{Troncoso:2019}.

A recent work of our group showed by means of micromagnetic simulations the possibility of exciting high-frequency ($\sim$ THz) propagating AFM modes in FiM strips~\cite{Sanchez-Tejerina:24}. Such micromagnetic modeling, based on the interaction between two antiparallel coupled magnetic sublattices, predicted the existence of forward volume spin waves (FVSWs) in strips of these materials with perpendicular uniaxial anisotropy, so that the main component of the net magnetization results perpendicular to the direction of SW propagation. 
Thus, in the case of spin angular momentum compensation between the two sublattices a dispersion curve was obtained that does not obey in general a quadratic dependence on the wave number $k$, but a more complex form that could only be fitted by adding successive even exponent terms to the fitting function. Outside the angular momentum compensation condition, the dispersion relation splits into two curves maintaining an intricate dependence on $k$, whose separation was strongly influenced by the strength of the coupling between sublattices.

In this work, we deal with the obtaining of theoretical expressions to account for the dispersion curves found for these complex systems near the AMC condition. To contrast our theoretical predictions, a set of micromagnetic simulations have also been carried out, showing an excellent match between theory and the numerical results, then unveiling the origin of the mentioned complex behavior of these coupled systems. 

To define the context in which the theoretical expressions of the dispersion curves of these FVSWs have been obtained, it must be established that FiMs can be described as formed by two strongly coupled sub-lattices through the exchange interaction, each sub-lattice represented by respective magnetizations that are aligned along the directions of the unit vectors $\vec{m}_i\ \left(i=1,2\right)$ at each mesoscopic point. The energy functional includes, in principle, symmetric exchange, Zeeman, magnetostatic, and anisotropy interactions. Antisymmetric exchange in the form of the Dzyaloshinkii-Moriya interaction can also be considered, but our calculations show that it is irrelevant for the modes here put on display. Additionally, no external field will be applied and magnetostatic effects will also be neglected, because of the relatively low value of the net magnetization. Accordingly, exchange interactions are expressed as
\begin{equation}
\varepsilon_{exch}=A_{ii}||\nabla\vec{m}_1||^2+A_{ii}||\nabla\vec{m}_2||^2+A_{ij}\langle\nabla\vec{m}_1,\nabla\vec{m}_2\rangle-B_{ij}\vec{m}_1\vec{m}_2\text{,}
\end{equation}
where $||M||$ represents the Frobenius norm of matrix $M$ and $\langle M_1 , M_2\rangle$ is the Frobenius inner product of matrices $M_1$ and $M_2$. $A_{ii}$ are the non-local intra-lattice exchange coefficients, while $A_{12}$ is the non-local inter-lattice exchange coefficient. As can be noted, the intralattice exchange parameter is assumed to be equal for both sublattices. Finally, $B_{12}$ determines the strength of the local inter-lattice exchange\cite{Sanchez-Tejerina:20}. Regarding the uniaxial anisotropy interaction, the expression  
\begin{equation}
\varepsilon_{anis}=-K_{u}\left(\vec{m}_1\cdot\vec{u}\right)^2-K_{u}\left(\vec{m}_2\cdot\vec{u}\right)^2
\end{equation}
is to be considered, $\vec{u}$ being the unit vector in the direction of the system easy axis ($K_{u}>0$). Again, the anisotropy constant has been assumed to be the same for both sublattices. The total energy density $\varepsilon$ is written as the sum of $\varepsilon_{exch}$ and $\varepsilon_{anis}$.

Magnetization dynamics is determined by two strongly coupled Landau-Lifshitz-Gilbert (LLG) equations~\cite{Puliafito:19,Martinez:20,Sanchez-Tejerina:24}
\begin{equation}\label{eq:LLG}
\dot{\vec{m}}_i=-\gamma_i\vec{m}_i\times\vec{B}_{\text{eff},i}+\alpha_i\vec{m}_i\times\dot{\vec{m}}_i+\vec{\tau}_{\text{SC},i}\qquad \left(i=1,2\right)\text{,}   
\end{equation}
where $\gamma_i$ and $\alpha_i$ are the gyromagnetic factor, and the damping parameter for each sub-lattice, respectively. $\gamma_i$ is proportional to the free-electron gyromagnetic factor $\gamma_e$ in the form $\gamma_i=\frac{g_i}{2}\gamma_e$, with $g_i$ being the corresponding Land\'{e} factor for each sublattice. $\vec{H}_{\text{eff},i}$ is the effective field, calculated as $\vec{B}_{\text{eff},i}=-\frac{1}{M_{s,i}}\frac{\delta\varepsilon}{\delta \vec{m}_i}$, and $\vec{\tau}_{\text{SC},i}$ corresponds to the torque associated with the injection of angular momentum (AMI) due to spin currents (SC). The right-hand side of (\ref{eq:LLG}) then includes two complementary terms: Gilbert damping and that AMI, i.e., angular momentum interchange with subsystems different from the magnetic system formed by the spins lattice. These subsystems are usually conduction electrons or the mechanical degrees of freedom of the crystal lattice (such as lattice vibrations). Since the aim of this work is to compute the excitation modes of the magnetic system, detailing either how the system is excited or how it is damped is not relevant, so these terms can be ignored or assumed to compensate for each other. Harmonic excitation will also be assumed to obtain periodic solutions in the direction of propagation of the SWs. Finally, since the study focuses on narrow (one-dimensional) strips, the propagation direction will be taken along the direction of the strip, which can be assigned to the $x$-axis.

Under these assumptions, and for low excitation amplitudes, the respective magnetizations for each sublattice are expected to vary in time and space in the form $\vec{m}_i =\vec{m}_{i,eq}+ \delta\vec{m}_i\mathrm{e}^{i\left(\omega t-kx\right)}$ , where $\vec{m}_{i,eq}$ represents the equilibrium state for the magnetization, perpendicular to the strip, and $\delta\vec{m}_i$ is a small in-plane vector that describes the perturbation from equilibrium. The appendix contains details of the calculation of the dispersion curves that relate the wavenumber $k$ to the frequency of excitation $\omega$. 
Two final expressions, constituting the main result of the work, are obtained in the form:
\begin{equation}\label{eq:dispersion}
    \omega_{\pm} = \frac{1}{2}\sqrt{\Gamma_A \Gamma_B\left(\nu_+^2-\nu_-^2\right)+\frac{1}{2}\left(\Gamma_A+\Gamma_B\right)^2\nu_-^2\left[1\pm\sqrt{\left(\frac{\Gamma_A-\Gamma_B}{\Gamma_A+\Gamma_B}\right)^2 +\frac{4\nu_+^2\Gamma_A\Gamma_B}{\nu_-^2\left(\Gamma_A+\Gamma_B\right)^2}}\right]}
\end{equation}
that establishes that two frequencies $f_\pm=\omega_\pm/\left(2\pi\right)$ are possible for each $k$, since $\Gamma_A$ and $\Gamma_B$ are both functions of the wavenumber in the form
\begin{subequations}\label{eq:definiciones}
\begin{align}
   \Gamma_A\left(k\right)&=2K_{u}+\left(2A_{ii}-A_{ij}\right)k^2\text{,}\\
    \Gamma_B\left(k\right)&=2\left(K_u-B_{ij}\right) +\left(2A_{ii}+A_{ij}\right)k^2\text{.}
\end{align}    
\end{subequations}
Here, the values $\nu_\pm$ are given by $\nu_\pm=\frac{1}{S_2}\pm\frac{1}{S_1}$, $S_i$ being the angular momenta of the sublattices. As a result, the values of $f_{\pm}$ converge under AMC recovering the expression for the uniaxial AFM materials~\cite{NotaRezende}, $\omega=\frac{\sqrt{\Gamma_A\Gamma_B}}{S}$. This condition can be achieved either at a certain temperature for a given composition or for a certain composition at a given temperature. Additionally, the uniaxial FM dispersion relation is recovered~\cite{Kalinikos:86} from Eq.~\ref{eq:dispersion} by considering equal angular momenta for both sublattices and setting to zero the interlattice exchange parameters, $A_{ij}$ and $B_{ij}$.%

\begin{figure*}
\centering
\includegraphics[width=\columnwidth]{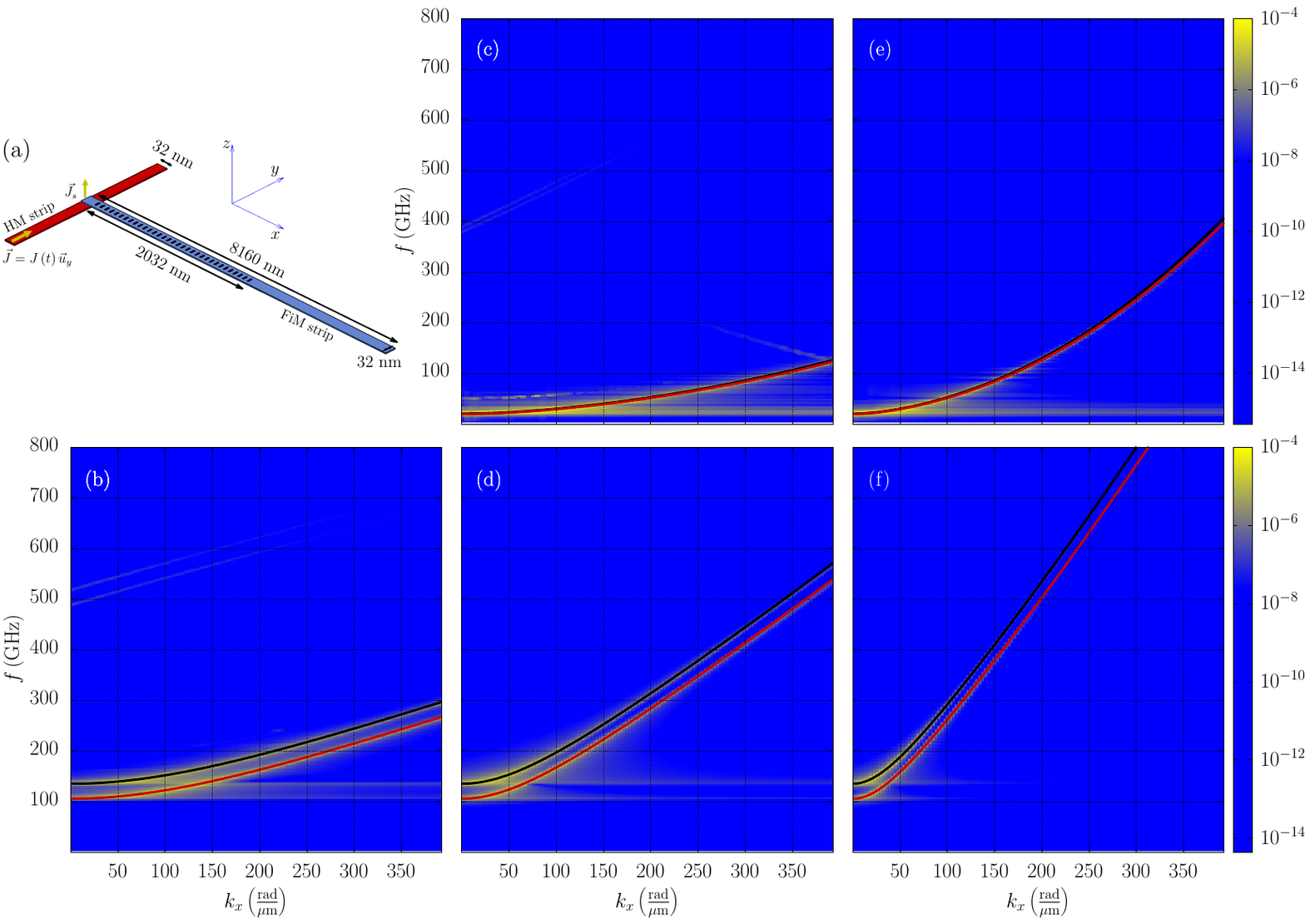}
\caption{\label{fig:01} (a) Sketch of the modeled device. (b)-(f) Dispersion diagrams of a FiM material obtained from $n_x$ with sublattices angular momenta $S_1 = 1.025\:S_2=34.04~\unit{\micro \joule.\second.\meter^{-3}}$ for various intra- and intersublattices exchanges: (b) $A_{ii}=0.1~\unit{fJ.m^{-1}}$, $A_{ij}=-6.0~\unit{pJ.m^{-1}}$, and $B_{ij}=-40~\unit{MJ.m^{-3}}$ (c) $A_{ii}=10~\unit{pJ.m^{-1}}$, $A_{ij}=-150~\unit{fJ.m^{-1}}$, and $B_{ij}=-1.0~\unit{MJ.m^{-3}}$, (d) $A_{ii}=10~\unit{pJ.m^{-1}}$, $A_{ij}=-6.0~\unit{pJ.m^{-1}}$, and $B_{ij}=-40~\unit{MJ.m^{-3}}$, (e) $A_{ii}=40~\unit{pJ.m^{-1}}$, $A_{ij}=-150~\unit{fJ.m^{-1}}$, and $B_{ij}=-1~\unit{MJ.m^{-3}}$ , (f) $A_{ii}=40~\unit{pJ.m^{-1}}$, $A_{ij}=-6.0~\unit{pJ.m^{-1}}$, and $B_{ij}=-40~\unit{MJ.m^{-3}}$. Colormaps are obtained from full micromagnetic simulations while solid lines corresponds to Eq.~\ref{eq:dispersion}.}
\end{figure*}

To prove the veracity of this result, we have proceeded to perform a series of numerical simulations where the FiM is assumed to be formed by computational cells, in a finite difference scheme, each cell containing two magnetization vectors, one for each sub-lattice, and both sub-lattices are coupled by an interlattice exchange interaction (see details in Ref.~\cite{Martinez:20b}). Simulations of the modeled device (sketched in Fig.~\ref{fig:01}(a)) have been carried out by our code~\cite{Martinez:19, Martinez:20, Martinez:20b} implemented on graphic processing units.
We note that Eq.~\ref{eq:dispersion} shows that the dispersion relation is determined by the angular momenta ratio, anisotropy, and intra- and interlattice exchange parameters. Additionally, saturation magnetization for each sub-lattice can be adjusted varying the temperature and/or composition. Thus, this latter parameter is employed to define different angular momenta ratios near the compensation condition:
\renewcommand{\theenumi}{(\roman{enumi}}%
\renewcommand{\labelenumi}{\theenumi)}%
\begin{enumerate}
    \item $M_{s,1} = 1.048\:M_{s,2}=1.094~\unit{MA.m^{-1}}$ and $S_1 = 1.075\:S_2=39.08~\unit{\micro \joule.\second.\meter^{-3}}$
    \item $M_{s,1} = 1.000\:M_{s,2}=0.953~\unit{MA.m^{-1}}$ and $S_1 = 1.025\:S_2=34.04~\unit{\micro \joule.\second.\meter^{-3}}$
    \item $M_{s,1} = 0.976\:M_{s,2}=0.887~\unit{MA.m^{-1}}$ and $S_1 = 1.000\:S_2=31.67~\unit{\micro \joule.\second.\meter^{-3}}$
    \item $M_{s,1} = 0.918\:M_{s,2}=0.742~\unit{MA.m^{-1}}$ and $S_1 = 0.941\:S_2=26.51\:\unit{\micro \joule.\second.\meter^{-3}}$.
\end{enumerate}
Note that the magnetization compensation condition (ii) differs from the angular momentum compensation condition (iii) due to distinct Land\'{e} factors in each sublattice, $g_1=2$, $g_2=2.05$. Other parameters have been chosen to be the same for both sublattices, as the non-local intralattice exchange parameter which is varied between $2\:\mathrm{pJ/m}$ and $40\:\mathrm{pJ/m}$, and the local interlattice exchange parameter which ranges from $-1\:\mathrm{MJ/m}^3$ to $-40\:\mathrm{MJ/m}^3$. It should be pointed out that the local and non-local interlattice exchange are not independent as both relate with the same exchange integral. However, this relation depends on the lattice constant and the number of other sublattice neighbors, which differs for different crystals. Therefore, for each local interlattice exchange parameter we perform three simulations with different non-local exchange parameters. For large values of interlattice exchange parameter, it is possible to stabilize a uniform strip with virtually zero intralattice exchange ($A_{ii}=0.1~\unit{fJ.m^{-1}}$). For those special cases, simulations with the latter intralattice exchange parameter have also been performed. 

The rest of significant parameters are those that can be found in the literature for a prototypical FiM as GdFeCo~\cite{Caretta:18}: anisotropy constant $K_{u,i}=0.1 \frac{\text{MJ}}{\text{m}^3}$ (refs. \cite{Joo:21,Kim:19,Sala:22,Kato:08}), spin-Hall angle $\theta_{SH}=0.15$ \cite{Martinez:20}, and damping constant $\alpha_i=0.001$. %

The device consists of a FiM strip and a current line of a heavy metal (HM) that runs perpendicular to the strip at one end, as shown in Fig.~\ref{fig:01}(b). The FiM material is initially uniformly magnetized in the out-of-plane direction, $z$. The FiM strip area is $8192~\text{nm}\times 32~\text{nm}$, and is 6 nm thick, discretized in $1\;\mathrm{nm}\times1\;\mathrm{nm}\times6\;\mathrm{nm}$ cells, while the HM strip is 32 nm wide. Consequently, the $32\;\text{nm}\times 32\;\text{nm}$ leftmost end surface of the FiM strip constitutes the excitation region of the device due to the SC generated via SOT from the HM. This SC is polarized along the $\pm\vec{u}_x$ direction depending on the instantaneous direction of the electric current along the HM strip. To excite the AFM strip, an (unnormalized) sinc-shaped electric current pulse has been used, $J\left(t\right)=J_0\mathrm{sinc}\left[2\pi f_c\left(t-t_c\right)\right]$, where  $J_0=1~\unit{\tera\ampere.m^{-2}}$ is the pulse amplitude, $t_c$ is a delay time, which has been chosen as one half of the total simulation time, set to $1~\text{ns}$ and $f_c$ is the excitation cut-off frequency, which was set to $1~\text{THz}$. This excitation configuration allow to probe all frequencies up to $1~\text{THz}$ with a resolution of $1~\text{GHz}$. Using this activation, we ensure that each mode is equivalently fed. In addition, this current is sufficiently small to remain in the linear regime of activation of damped, stable oscillations\cite{Cheng:16} and to avoid any static changes in the AFM phases of the sample. The delay time also provides a reasonable offset to the peak of the pulse, allowing a gradual increase in amplitude from the beginning of the simulation. To monitor the value of the local N\'eel vector $\vec{n}=\vec{m}_1-\vec{m}_2$, a set of 254 regularly spaced probes is located on the FiM strip. The probes are $3\;\text{nm}\times 8\;\text{nm}$ in size and 5 nm apart from each other, then occupying a total length of $2032\;\text{nm}$. 

\begin{figure}
\centering
\includegraphics[width=.5\columnwidth]{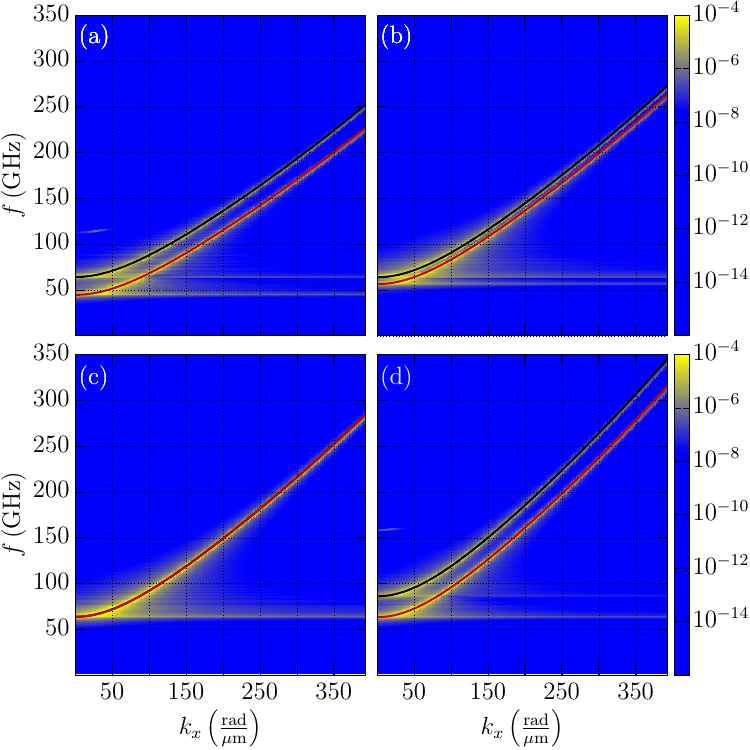}
\caption{\label{fig:02} Dispersion diagrams of a FiM material with exchange parameters: $A_{ii}=10~\unit{pJ.m^{-1}}$, $A_{ij}=-1.5~\unit{pJ.m^{-1}}$, and $B_{ij}=-10~\unit{MJ.m^{-3}}$ for different sublattices angular momenta relations: (a) $S_1 = 1.075\:S_2=39.08~\unit{\micro \joule.\second.\meter^{-3}}$ (i), (b) $S_1 = 1.025\:S_2=34.04~\unit{\micro \joule.\second.\meter^{-3}}$ (ii), (c) $S_1 = 1.000\:S_2=31.67~\unit{\micro \joule.\second.\meter^{-3}}$ (iii), (d) $S_1 = 0.941\:S_2=26.51\:\unit{\micro \joule.\second.\meter^{-3}}$ (iv). Color maps are obtained from full micromagnetic simulations while solid lines correspond to Eq.~\ref{eq:dispersion}.}
\end{figure}

To check the analytical expression given by Eq.~\ref{eq:dispersion} we first fixed the angular momenta ratio to Scenario (ii) $M_{s,1} = 1.000\:M_{s,2}=0.953~\unit{MA.m^{-1}}$ and $S_1 = 1.025\:S_2=34.04~\unit{\micro \joule.\second.\meter^{-3}}$ corresponding to the magnetization compensation point. This allows us to clearly distinguish the two dispersion curves predicted from our calculations. We then study the effect of the different intra- and inter-lattice exchange parameters. Figure~\ref{fig:01}(b)-(f) depict the cases with lowest and largest intra- and interlattice exchange parameters in the considered range, as well as a case with a middle value of the intralattice exchange parameter. All combinations of parameters show an excellent agreement between Eq.~\ref{eq:dispersion} and full micromagnetic simulations in the whole range of frequencies and wave numbers. As predicted from the analytical expression, the effect of local interlattice exchange is double. First, a larger local interlattice exchange increases the frequency of the uniform precession, representing the minimum available frequency. This is translated into a shift of the whole branch towards higher frequencies, as can be checked by comparing Fig.~\ref{fig:01}(c) and (e) with Fig.~\ref{fig:01}(b),(d) and (f), thus increasing the phase velocity. In addition, the frequency difference between the two branches also increases for larger local interlattice exchanges. Moreover, as should be expected, non-local exchange, neither intralattice nor interlattice, affects the uniform mode as can be checked comparing Fig.~\ref{fig:01}(b) with Fig.~\ref{fig:01}(d) or (f) and Fig.~\ref{fig:01}(c) with Fig.~\ref{fig:01}(e). Nevertheless, those terms greatly affect the group velocity, being larger for larger non-local exchange parameter. 

To further corroborate the validity of Eq.~\ref{eq:dispersion} we now fix the intra- and interlattice exchange parameters to $A_{ii}=10~\unit{pJ.m^{-1}}$, $A_{ij}=-1.5~\unit{pJ.m^{-1}}$, and $B_{ij}=-10~\unit{MJ.m^{-3}}$, and vary the angular momenta ratio. Figure~\ref{fig:02} represents the dispersion relations for the four different specified scenarios, (i) $S_1 = 1.075\:S_2=39.08~\unit{\micro \joule.\second.\meter^{-3}}$, (ii) $S_1 = 1.025\:S_2=34.04~\unit{\micro \joule.\second.\meter^{-3}}$, (iii) $S_1 = 1.000\:S_2=31.67~\unit{\micro \joule.\second.\meter^{-3}}$, and (iv) $S_1 = 0.941\:S_2=26.51\:\unit{\micro \joule.\second.\meter^{-3}}$. As expected, at the AMC condition (iii), the two branches merge to a single one, and the larger the difference between the two sublattices angular momentum density, the larger the frequency shift between the two branches for a fixed wave number. In addition, the lower the angular momentum densities, the higher frequencies are reached. Nevertheless, we note that we have kept the energy densities constant. Finally, it should be noted that in all cases the agreement between Eq.~\ref{eq:dispersion} and the full micromagnetic simulations is excellent.

In conclusion, we have tested the analytical expression of the dispersion curves of FVSW in FiMs against numerical simulations of these materials, then revealing the dependencies of these curves on the different parameters governing magnetic dynamics in them. Since FiMs are considered as formed by two strongly coupled magnetic sublattices, there is a relevant role played by the inter-lattice exchange interactions, that impede from independently describing the behavior of each sub-lattice. Two dispersion relations are worked out, that converge as the system approach to the AMC condition. These analytical expressions quantify the role played by the uncompensation of the two sublattices near the AMC condition, giving a better insight on the underlying physical phenomena. In all cases, the work confirms the possibility of exciting high-frequency, close to the THz band, propagating AFM modes in FiM strips. Equation~\ref{eq:dispersion} also assists the choice of materials and composition/temperature to excite SWs with specific frequency and wavelength. Accordingly, the work facilitates the design of technological applications in this frequency range based on FiMs. Specifically, this makes it possible to implement logic devices based on SWs, as it will be the subject of a forthcoming work.

\begin{acknowledgments}
We acknowledge funding from Ministerio de Ciencia, Innovaci\'on y Universidades of the Spanish Government through Project PID2023-150853NB-C31.
\end{acknowledgments}

\bibliography{mmag}% 

\begin{thebibliography}{30}
\bibitem [{Not()}]{NotaRezende}%
  \BibitemOpen
  \href@noop {} {}\bibinfo {note} {See for example eq. (36b)
  from~\cite{Rezende:19}, taking into account that the micromagnteic approach
  is limit to low $k$ and that expression is limit to small
  anisotropy}\BibitemShut {NoStop}%
\end{thebibliography}

\appendix*

\section{calculation of the dispersion relation}
This appendix deals with the calculation of the dispersion curves for the propagation of FVSWs in FiMs with uniaxial anisotropy. Calculations are made around the AMC point for small excitation amplitudes. As stated in the main text, the approach of small amplitudes allows us to decompose the magnetization as a sum of an equilibrium value plus a small harmonic value in time and space, in the form 
\begin{equation}\label{eq:senpeq}
\vec{m}_{i}=\vec{m}_{i,eq}+\delta\vec{m}_i e^{i\left(\omega t - kx\right)}\text{.}
\end{equation}
This expression is to be applied to the resultant effective field $\vec{B}_{\text{eff},i}=-\frac{1}{M_{s,i}}\frac{\delta\varepsilon}{\delta \vec{m}_i}$, where, for the sake of completeness, $\varepsilon$ also includes the Dzyaloshinkii-Moriya interaction (DMI)\cite{Martinez:19,Sanchez-Tejerina:20}

\begin{equation}
\begin{split}
\varepsilon_{dmi}&=D_{1} \left( \left( \vec{m}_{1}\cdot\vec{u}_n \right) \nabla \cdot \vec{m}_1 - \vec{m}_1 \cdot \nabla\left( \vec{m}_{1}\cdot\vec{u}_n \right) \right) +\\
&+D_{2}\left( \left( \vec{m}_{2} \cdot \vec{u}_n \right) \nabla \cdot \vec{m}_2 - \vec{m}_2 \cdot \nabla\left( \vec{m}_{2}\cdot\vec{u}_n \right) \right) \text{,}
\end{split}
\end{equation}

although we will later prove that this interaction does not affect the dispersion relations for FVSW under the assumptions stated above.  Likewise, the effective field can be written as a sum of an equilibrium value plus a small harmonic value, that is,
\begin{equation}
\vec{B}_{eff,i}=\vec{B}_{i,eq}+\vec{B}_{i,dyn}\left(t,x\right)\text{.}
\end{equation}

Therefore, Eq.~\ref{eq:LLG} of the main text can be written:

\begin{equation}
i\omega \delta\vec{m}_i e^{i\left( \omega t- kx \right)} =-\gamma_i \left( \vec{m}_{i,eq}\times\vec{B}_{i,dyn} \right)-\gamma_i \left( \delta\vec{m}_{i}\times\vec{B}_{i,eq}+\delta\vec{m}_{i}\times\vec{B}_{i,dyn} \right)e^{i\left( \omega t- kx \right)}\text{,}
\end{equation}
where the damping and antidamping term have been omitted. It should be noted that the addend $\vec{m}_{i,eq}\times\vec{B}_{i,eq}=0$ as it is the torque at equilibrium. We can further simplify the expression by considering that the effective field at equilibrium is directed along $\pm\vec{u}_z$ and for small enough excitations we can assume $\delta\vec{m}_i \perp \vec{u}_z$. Consequently, Eq.~\ref{eq:LLG} reduces to

\begin{equation}
\begin{split}
i\omega \delta\vec{m}_i =&\pm\gamma_i \left( \left( \frac{B_{ij}}{M_{S,i}} - \frac{A_{ij}}{M_{S,i}}k^2 \right)\delta\vec{m}_j - \frac{2A_{ii}}{M_{S,i}}k^2\delta\vec{m}_i  \right)  \times\vec{u}_{z}\pm
\\ &\pm\gamma_i\left(\left(\frac{B_{ij}}{M_{S,i}}-\frac{2K_u}{M_{S,i}}\right) \delta\vec{m}_{i} \right)\times\vec{u}_z \text{,}\label{eq:A4}
\end{split}
\end{equation}
after the second order terms in $\delta\vec{m}_i$ have been neglected. It is worth noting that, under these assumptions, the terms depending on the Dzyaloshinkii-Moriya interaction (DMI) no longer play any role in the dynamics, because either they are second-order terms or dependent on the scalar product of $\delta\vec{m}_i$ and $\vec{u}_z$, vectors which have been assumed to be perpendicular.

Considering the angular momentum densities $S_i=\frac{M_{S,i}}{\gamma_i}$, and after some algebra, we get

\begin{equation}
i\omega S_i \delta\vec{m}_i =\mp \left[  \left( 2K_u-B_{ij} + 2A_{ii}k^2 \right)\delta\vec{m}_i - \left( B_{ij} - A_{ij}k^2 \right)\delta\vec{m}_j \right]  \times\vec{u}_{z} \text{.}\label{eq:A5}
\end{equation}

We can now recast this last result in terms of small magnetization $\vec{m}_+=\vec{m}_1+\vec{m}_2$ and N\'eel vector $\vec{l}=\vec{m}_1-\vec{m}_2$, whose combined dynamic components can account for the dynamic components of $\delta\vec{m}_1$ and $\delta\vec{m}_2$ in the form $\delta\vec{m}_1=\dfrac{\delta\vec{m}_{+}+\delta\vec{l}}{2}$ and $\delta\vec{m}_2=\dfrac{ \delta\vec{m}_{+}-\delta\vec{l}}{2}$. Introducing $\nu_+ = \frac{S_1+S_2}{S_1 S_2}$ and $\nu_- = \frac{S_1-S_2}{S_1 S_2}$ as angular momentum decompensation factors, Eq.\ref{eq:A5} transforms into the following couple of equations:

\begin{subequations}
\begin{align}
\begin{split}
i2\omega \delta\vec{m}_{+} &=\left\{ \nu_- \left[ 2K_u-2B_{ij} + \left(2A_{ii} +A_{ij} \right)k^2 \right]\delta\vec{m}_{+}-\right.\\
&\qquad\left.-\nu_+ \left[ 2K_u + \left(2A_{ii} -A_{ij}\right)k^2\right] \delta\vec{l} \right\}  \times\vec{u}_{z}\text{,} 
\end{split}\\
\begin{split}
i2\omega \delta\vec{l} &=\left\{ - \nu_+\left[ 2K_u-2B_{ij} + \left(2A_{ii} +A_{ij} \right)k^2 \right] \delta\vec{m}_{+} +\right.\\
&\qquad\left.+\nu_- \left[ 2K_u + \left(2A_{ii} -A_{ij}\right)k^2\right] \delta\vec{l} \right\}  \times\vec{u}_{z}\text{,}
\end{split}
\end{align}
\end{subequations} 
which is a set of coupled equations for the components of $\delta\vec{m}_+=\left(\delta m_{+,x},\delta m_{+,y}\right)$ and $\delta\vec{l}=\left(\delta l_x,\delta l_y\right)$ in the form

\begin{subequations}
\begin{align}
\left(4\omega^2 - \nu_-^2 A  - \nu_+^2 B \right) \delta m_{+,x} &= - \nu_+\nu_- \left( B + C \right) \delta l_x\text{,}\\
\left(4\omega^2 - \nu_-^2 A - \nu_+^2 B \right) \delta m_{+,y} &= - \nu_+\nu_- \left( B + C \right) \delta l_y\text{,}\\
\left(4\omega^2 - \nu_+^2 B - \nu_-^2 C \right) \delta l_x &=  - \nu_+\nu_- \left( A + B \right) \delta m_{+,x}\text{,} \\ 
\left(4\omega^2 - \nu_+^2 B - \nu_-^2 C \right) \delta l_y &= - \nu_+\nu_- \left( A+ B \right) \delta m_{+,y}\text{,}
\end{align}
\end{subequations}
where we have introduced the parameters
\begin{subequations}
\begin{align}
A &=  \left[ 2\left(K_u-B_{ij}\right) + \left(2A_{ii} +A_{ij} \right)k^2 \right]^2\text{,} \\
B &=  \left[ 2K_u + \left(2A_{ii} -A_{ij}\right)k^2\right] \left[ 2\left(K_u-B_{ij}\right) + \left(2A_{ii} +A_{ij} \right)k^2 \right]\text{,}\\
C &=  \left[ 2K_u + \left(2A_{ii} -A_{ij}\right)k^2\right]^2\text{.}
\end{align}\label{eq:defpar1}
\end{subequations}

The decoupling of the equations results in a new set of equations all of them leading to the same biquadratic equation in $\omega$
\begin{equation}
\omega^4-\left( 2\nu_+^2 B+ \nu_-^2\left( A+C \right) \right)\frac{\omega^2}{4}+\frac{1}{16}\left( \nu_+^4 B^2 +\nu_-^4 AC- \nu_+^2\nu_-^2 \left(AC+B^2\right) \right)=0\text{.}\label{eq:biquadratic}
\end{equation}

whose solutions are those given in the main text as Eq.~\ref{eq:dispersion} when the definitions in Eq.~\ref{eq:definiciones} of the main text are considered.

\end{document}